\documentclass[aps,twocolumn,showpacs,prl]{revtex4}
\usepackage{amssymb}
\usepackage{natbib}
\usepackage{amsmath}
\usepackage{amsfonts}
\usepackage{graphicx}
\usepackage{mathrsfs}
\usepackage{dcolumn}
\usepackage{bm}
\usepackage{color}
\usepackage{cancel}
\usepackage{mathbbol}
\usepackage{epsfig}
\usepackage{units}

\definecolor{rot}{rgb}{0.75,0.05,0.25}
\definecolor{hellgrau}{gray}{0.5}
\definecolor{blau}{rgb}{0,0,0.7}

\begin{document}

\title{Logarithmic oscillators: ideal Hamiltonian thermostats}
\author{Michele Campisi, Fei Zhan, Peter Talkner, and Peter H\"anggi}
\affiliation{Institute of Physics, University of Augsburg,
  Universit\"atsstr. 1, D-86135 Augsburg, Germany}
\date{\today }

\begin{abstract}
A logarithmic oscillator (in short, log-oscillator) behaves like an
ideal thermostat because of its infinite heat capacity: when it weakly couples to another system,
 time averages of the system observables agree with ensemble
 averages from a Gibbs distribution with a temperature $T$
 that is given by the strength of the logarithmic potential.
The resulting equations of motion are Hamiltonian and  may be
implemented not only in a computer but also with real-world experiments,
e.g., with cold atoms.
\end{abstract}


 \maketitle

Thermostats play an important role in computational physics \cite{Klages}.
They provide effective and useful methods to simulate the action of a thermal
environment on systems of physical and chemical interest. Mathematically speaking,
their salient feature is to produce ``thermostated dynamics'' of the system of
interest: that is,  they are meant to impose long-time averages of system observables
that coincide with  Gibbs-ensemble averages at a given temperature $T$.
Widely used thermostats are: the  Langevin thermostat \cite{Ermak80JCmptP169},
Andersen's stochastic collision thermostat  \cite{Andersen80JCP72} and the
Nos\'e-Hoover deterministic thermostat \cite{Nose84JCP81,Hoover85PRA31,Martyna92HCP97}.

Here we present a thermostat differing in various respects from the previously reported ones.
Our main result is that a
\emph{logarithmic oscillator} (or a ``log-oscillator'' as we shall call it
below), weakly coupled to the system of interest (in short ``the system''
in what follows) leads to thermostated system dynamics.
In its simplest 1D version the system+log-oscillator Hamiltonian reads:
\begin{equation}
 H= \sum_i \frac{p_i^2}{2m_i} + V(\mathbf q) +
\frac{P^2}{2M}+T\ln\frac{|X|}{b}+ h(\mathbf q,X)
\label{eq:H}
\end{equation}
where $\mathbf{p}=(p_1, \dots, p_N),\mathbf{q}=(q_1, \dots, p_N),m_i$
are the momenta, positions, and masses of the particles composing the
system; $X,P,M$ are the log-oscillator position, momentum, and mass,
respectively; $b>0$ sets the length scale of the log-oscillator and
$T$ is the thermostat temperature; $V(\mathbf{q})$ is the system inter-particle
 potential and $h(\mathbf q,X)$
denotes a {\it weak} interaction energy  that couples the log-oscillator to
the system. When the total Hamiltonian $H$ is ergodic, the system+log-oscillator 
trajectory samples the microcanonical ensemble, and the system trajectory samples
the canonical ensemble at temperature $T$.
This continues to hold if the 1D log-oscillator is replaced by higher dimensional 
log-oscillators, for example for a charged particle in the attractive logarithmic 
2D Coulomb field generated by a long charged wire.

\begin{figure}[b]
\includegraphics[width= 0.5 \linewidth]{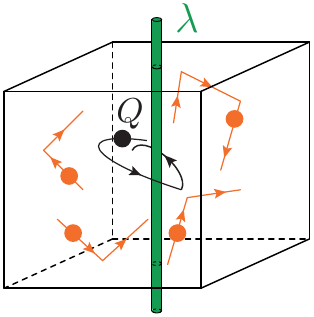}
\caption{(Color online) A single ion (black sphere) carrying the charge $Q>0$, 
subject to the attractive 2D Coulomb potential generated by a wire
(green cylinder) carrying the linear charge density $\lambda < 0$,
thermalizes, by means of short-range
repulsive collisions, a gas of neutral atoms (orange spheres) to the Gibbs distribution
of temperature $T=|Q \lambda|/\pi \varepsilon_0$.
}
\label{fig:sketch}
\end{figure}

Compared to the previously reported thermostats the present
thermostat exhibits an evident advantage. The Hamiltonian (\ref{eq:H}) 
or its higher dimensional versions can be
readily implemented in a physical experiment. 
In Fig. \ref{fig:sketch} we show a possible implementation.
The system is composed of a gas of neutral atoms confined
into a box. The thermostat is an ion subject to the attractive 2D coulomb potential
generated by a thin oppositely charged wire,
$|Q \lambda|/2 \pi \varepsilon_0  \ln \rho$.
Here $Q>0$ is the charge of
the ion, $\lambda<0$ the linear charge density of the wire,
$\rho$ the distance between wire and particle, and $\varepsilon_0$ the electric permittivity of vacuum.
Through short-range repulsive interactions
the ion thermalizes the neutral gas to the temperature 
$T=|Q \lambda|/\pi \varepsilon_0$.
Another possibility for the realization of a log-oscillator
is by means of a laser beam with an intensity profile of logarithmic form 
coupled non-resonantly to an atom \cite{Schleich10PRA82}.
This could be realized to thermostat cold atomic gases \cite{Bloch08RMP80}.

Atomic systems in isolation from the environment naturally sample the microcanonical 
ensemble. For small systems this sampling may considerably differ from the canonical one
and can result in distinctive thermodynamic features such as negative specific heats. 
These were experimentally investigated with
small atomic clusters \cite{Schmidt01PRL86,Gobet02PRL89}. Typically it is difficult to 
have a small isolated system sample the canonical Gibbs distribution.
Our method opens this possibility. More generally, 
by using a single log-oscillator as an environment simulator,
our method allows to experimentally study thermostated small systems
in isolation from the real environment. One advantage of this
paradigm is that our method would allow to control a thermal parameter,
the temperature $T$, by means of mechanical parameters, e.g., with reference 
to Fig. \ref{fig:sketch}, the charge density $\lambda$ on the wire.

Just like the Nos\'e-Hoover thermostat, our thermostat is deterministic and time-reversible,
but at variance with Nos\'e-Hoover dynamics which are not Hamiltonian  \cite{Klages,Kusnezov},
our thermostated dynamics are manifestly Hamiltonian. 
There exist ``generalized Hamiltonian formalisms'' \cite{Klages}
for the Nos\'e-Hoover dynamics in the literature. The most prominent examples
use Nos\'e's Hamiltonian \cite{Nose84JCP81}: $H_\text{Nos\'e}=\sum {{p}_i^2}/{2m_i {X}^2} + V({\mathbf{q}}) +
{{P}^2}/{2M}+f T\ln {{X}}$ or Dettmann's Hamiltonian \cite{Dettmann,Hoover12arXiv}: $H_D= X H_\text{Nos\'e} $.
At variance with our Hamiltonian in Eq. (\ref{eq:H}), these involve the non standard 
kinetic terms, ${{p}_i^2}/{2m_i {X}^2}$ and ${{p}_i^2}/{2m_i {X}}$,
respectively, which, due to the dependence on the log-oscillator position,
cannot readily be realized in an experiment.
Further, while Dettmann's Hamiltonian produces thermostated
trajectories only for a specific value of the energy (i.e., $H_D=0$) our method 
thermalizes the system irrespective of the energy value.
We elucidated these issues further in Ref. \cite{Campisi12arXiv}.
The usefulness and importance of the Nos\'e-Hoover equations as a computational
thermostat are beyond question \cite{HooverBook}.

\emph{Theory.--}
Before we shall provide the formal argument we present a physical explanation 
indicating why it is plausible that the Hamiltonian in Eq. (\ref{eq:H}) leads to thermostated 
system dynamics. Consider the isolated 1D log-oscillator:
\begin{equation}
H_{\text{log}} = \frac{P^2}{2M}+T\ln\frac{|X|}{b}\, .
\label{eq:Hlog}
\end{equation}
Applying the virial theorem,
$
 \left\langle
P \partial H_\text{log}/\partial P \right\rangle=
\left\langle
X \partial H_\text{log}/\partial X \right\rangle
$,
to the 1D log-oscillator, we obtain
$
\left\langle {P^2}/{M} \right\rangle = T
$
where $\langle \cdot \rangle$ denotes the time average.
That means that all trajectories of a log-oscillator have the same average
kinetic energy \cite{Schleich10PRA82}, i.e., the same kinetic
temperature $\langle P^2/M \rangle=T$, regardless of their energy $E$. This
implies $\partial T/\partial E=0$. Recalling the definition of heat capacity, $C=
\partial E/\partial T$, one finds that the log-oscillator exhibits a
spectacular property: its heat capacity is infinite, which is the defining
feature of an ideal thermostat. Since the log-oscillator may only exist in
the state of temperature $T$, we expect that a system will reach this
same temperature $T$  when it is weakly coupled to the log-oscillator.

To formally prove that the log-oscillator induces thermostated dynamics of the
system at the temperature $T$, we recall
the general expression for the probability density function
$p(\mathbf{q},\mathbf{p})$ to find a system at the point
$(\mathbf{q},\mathbf{p})$ of its phase space when it is weakly coupled
to a second system [the log-oscillator in the present case], provided that
the compound system probability distribution is microcanonical. It reads
\cite{Khinchin49Book}
\begin{eqnarray}
 p(\mathbf{q},\mathbf{p})=\frac{\Omega_ {\text{log}}[E_{\text{tot}}-H_S(\mathbf{q},
\mathbf{p})]}{
\Omega(E_{\text{tot}})}\, ,
\label{eq:khinchin-formula}
\end{eqnarray}
where $E_{\text{tot}}$ is the total (conserved) energy of
the compound system. With $E$ denoting the log-oscillator energy,
\begin{equation}
 \Omega_{\text{log}}(E) = \int \mathrm{d}{X}
\mathrm{d}{P}\delta[E-H_{\text{log}}(X, P)]
\label{eq:Omega-B}
\end{equation}
is the density of states of the log-oscillator, and
\begin{equation}
 \Omega(E_{\text{tot}}) = \int \mathrm{d}{X}
\mathrm{d}{P}\mathrm{d}\mathbf{q}\mathrm{d}\mathbf{p}\,
\delta[E_{\text{tot}}-H(\mathbf{q},\mathbf{p},X,P)]
\end{equation}
is the density of states of the compound system. Here $\delta(...)$ denotes
Dirac's delta function and $H_S$ is the system Hamiltonian.

According to Eq. (\ref{eq:khinchin-formula}) the density of states of the
log-oscillator defines the shape of the distribution of the system.
Performing the integration in Eq. (\ref{eq:Omega-B}) with the log-
oscillator Hamiltonian, Eq. (\ref{eq:Hlog}), one obtains for the density
of states of the log-oscillator the expression
\begin{equation}
\Omega_{\text{log}}(E)=  2 b\sqrt{2 \pi M/T}\,  e^{E/T}\, .
\label{eq:Omega-log}
\end{equation}
Inserting Eq. (\ref{eq:Omega-log}) into Eq. (\ref{eq:khinchin-formula}) yields the Gibbs
distribution for the system,
\begin{eqnarray}
 p(\mathbf{q},\mathbf{p})={e^{-H_S(\mathbf{q},\mathbf{p})/T}
}/{Z(T)} \, ,
\label{eq:canonical}
\end{eqnarray}
regardless of the energy $E_{\text{tot}}$ assigned to the compound system.
Here $Z(T)=\int \mathrm{d}\mathbf{q}\mathrm{d}\mathbf{p}e^{-
H_S(\mathbf{q},\mathbf{p})/T}$ is the system canonical partition
function.

Also a $f$-dimensional log-oscillator
$
H(\mathbf{X},\mathbf{P})=\mathbf{P}^2/(2M) +f T/2  \ln
(\mathbf{X}^2/b^2)$
[where $\mathbf{X}$ and $\mathbf{P}$ are vectors of size $f$]
results in the exponential density of states $\Omega_{\text{log}} \propto e^{E/T}$.
Therefore, $f$-dimensional log-oscillators induce thermostated
dynamics as well.

So far we have left the system-thermostat interaction $h(\mathbf{q},X)$ unspecified.
As in standard statistical mechanics where a heat bath with many degrees of
freedom replaces the single log-oscillator \cite{Khinchin49Book},
$h(\mathbf{q},X)$ must comply with two requirements: (i) it must be 
sufficiently weak that it can completely be neglected in the calculation of 
the probability density $p(\mathbf{q},\mathbf{p})$. 
This assumption guarantees the applicability of Eq. (\ref{eq:khinchin-formula})
provided that  the total system
stays in microcanonical equilibrium. In order that this equilibrium state actually is
reached from arbitrary initial conditions it is necessary (ii) that the total dynamics is
ergodic. To meet these two requirements, 
short-range repulsive interactions typically suffice, see the numerical examples below.
Note that with a short-range repulsive interaction, the fraction of time during which the 
log-oscillator interacts with any other particle is much smaller
than one. This assures that the average interaction energy represents only a small part
of the total energy, and hence the weak coupling assumption 
implied by Eq. (\ref{eq:khinchin-formula}) is met.

\emph{Numerics.--}
In order to corroborate our statement we performed 1D and
3D molecular dynamics simulations using symplectic integrators
\cite{Hairer06Book}.

\begin{figure}[b]
\includegraphics[width= \linewidth]{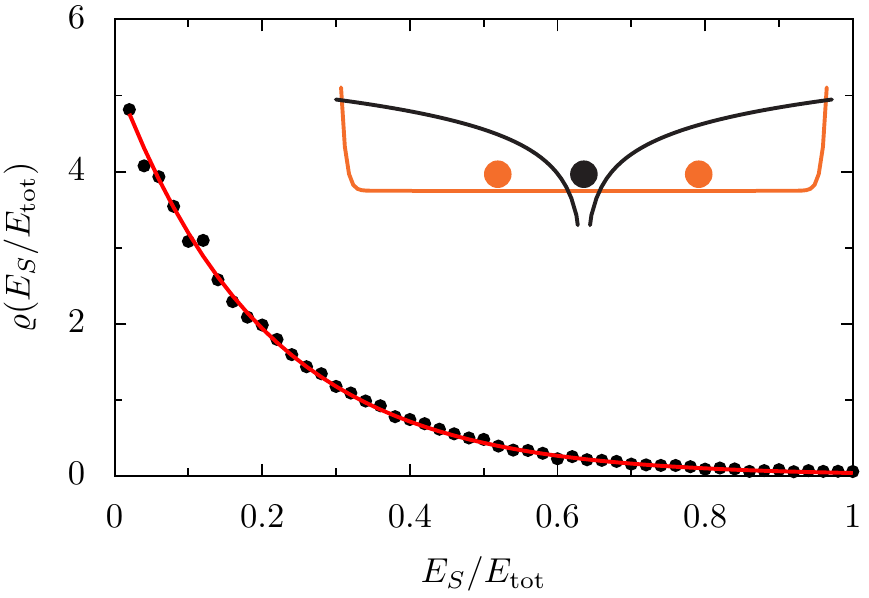}
\caption{(Color online)
Probability density function of energy for a
system of two particles in a 1D box performing short ranged
collisions with a log-oscillator.
The system energy $E_S$ is rescaled by the total simulation energy which is
$E_{\text{tot}}=75 \varepsilon$. The log-oscillator strength is $T=15 \varepsilon$
and the box length is $L=10 \,e^{E_{tot}/T}\sigma \simeq 1484 \sigma$.
Black dots: numerical simulation. Red line: Gibbs distribution at
temperature $T=15 \varepsilon$.
The total system is schematically represented
in the inset with the system of interest (two orange particles) confined to the box potential
(orange curve), and the log-oscillator (black particle) confined to the logarithmic potential
(black curve).
}
\label{fig:plot}
\end{figure}

In our first numerical experiment we used
two point particles of mass $m$ in a 1D box of length
$L$ and placed a log-oscillator of mass $M$ and
strength $T$ between them, see the inset in Fig. \ref{fig:plot}.
The three particles interact with each other and with the fixed walls via
the truncated Lennard-Jones potential, reading
\begin{equation}
V_{LJ}(q)=
 \left\{
  \begin{array}{ll}
0\, , &  |q| > 2^{1/6}\sigma \\
4\varepsilon \left[\left(\frac{\sigma}{q}\right)^{12}-\left(\frac{\sigma}{q}
\right)^ { 6 } \right ] +\varepsilon\, , & |q| < 2^{1/6}\sigma
 \end{array}
 \right. \, ,
\label{eq:V(x)}
\end{equation}
that is $h(q_1,q_2,X)=\sum_i V_{LJ}(|q_i-X|)$, and 
$V(q_1,q_2)=V_{LJ}(|q_1-q_2|)+\sum_i [V_{LJ}(|q_i+L/2|)+V_{LJ}(|q_i-L/2|)]$ where $L$ is the box length.
In the simulations we adopted $m,\sigma$ and $\varepsilon$, as the units of mass,
length, and energy, respectively.
In order to avoid the singularity of the logarithmic potential at the origin
we replaced it with the following potential:
 \begin{equation}
 \varphi_{b}(X)= \frac{T}{2}\ln\frac{X^2+b^2}{b^2} \, .
 \label{eq:phi(x)}
\end{equation}
For all simulations we used the value $b=\sigma$.
This truncation results in a correction of the density of states
(\ref{eq:Omega-log}), which  vanishes as the energy $E_{\text{tot}}$ increases.
Fig. \ref{fig:plot} depicts the probability density function, $\varrho(E_S)$
of finding the system consisting of the two orange particles depicted
in the inset at the kinetic energy  $E_S$  in a molecular dynamics simulation
at total energy $E_{\text{tot}}$.
According to Eq.  (\ref{eq:canonical}) this should be of the form
$\varrho(E_S)\propto e^{-E_S/T}\Omega_S(E_S)\propto e^{-E_S/T}$,
where  $\Omega_S(E_S)$ is the system density of states. Note that $\Omega_S(E_S)$ is
constant in the case of a system Hamiltonian composed of two quadratic degrees of
freedom.  The numerically computed curve excellently fits the desired canonical
distribution with the expected temperature $T$. The simulation energy
$E_{\text{tot}}$ was chosen large enough, so that the 
error introduced by the replacement of the purely logarithmic potential with
the truncated one, was negligible. The box length was
taken such that it exceeded the maximal excursion of the
log-oscillator $x_\text{max}=2 \sigma \sqrt{e^{2E_\text{tot}/T}-1}$. Otherwise the log-potential would be effectively cut-off by
the box-potential and consequently  the exponential shape of the
density of states would be destroyed.

\begin{figure}[b]
\includegraphics[width= \linewidth]{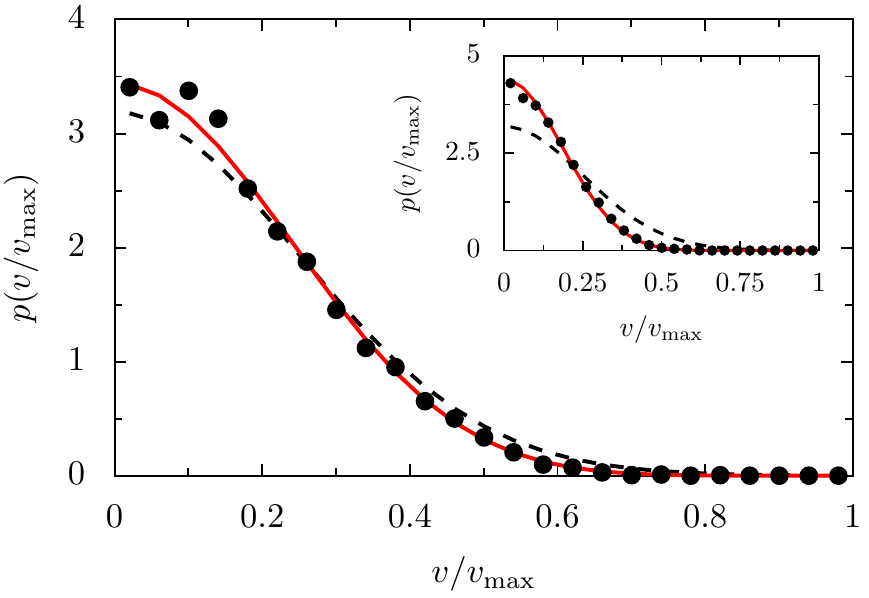}
\caption{(Color online) Probability density function of the
absolute value of the velocity components of a gas of
$3$ neutral particles in a confining box, weakly interacting with a charged particle
in a truncated 2D Coulomb field, Eq. (\ref{eq:H-coulomb2D-trunc}). The velocity is rescaled by
the maximum possible velocity
$v_{\text{max}}= (2 E_{\text{tot}}/m)^{\nicefrac{1}{2}}$. The simulation energy is
$E_{\text{tot}}=120 \varepsilon$, $T = 15 \varepsilon $, $b=\sigma$,
and the box has dimensions $L_z=20\sigma,L_x=L_y= 2 r_{\text{max}}$
($r_\text{max}= \sigma (e^{E_{\text{tot}}/T}-1)^{\nicefrac{1}{2}} \simeq 109 \sigma$ is the
maximal possible excursion of the log-oscillator).
Black dots: numerical simulation. Black dashed line: Maxwell distribution
at temperature $T= 15 \varepsilon $. Red solid line: corrected distribution accounting for the
truncation of the log-potential Eq. (\ref{eq:H-coulomb2D-trunc}).
Inset: same, but for a gas of 8 particles.
}
\label{fig:plot2}
\end{figure}
Our second numerical experiment considers as thermal bath
a charged particle in the electric field generated by a long
and oppositely charged wire: the so-called 2D Coulomb
potential which is of logarithmic form, Fig. \ref{fig:sketch}. 
The charged particle
Hamiltonian reads:
\begin{equation}
H(P_x,P_y,P_z,X,Y,Z)=\frac{P_x^2+P_y^2+P_z^2}{2M} +T  \ln
\frac{X^2+Y^2}{b^2}
\label{eq:H-coulomb2D}
\end{equation}
where $T=|Q \lambda|/\pi \varepsilon_0 $.
Assuming that the motion is confined in the $Z$ direction by two rigid
walls parallel to the $XY$ plane and separated by a distance $L_z$,
one obtains for the density of states the expression
$
\Omega_{\text{log}}(E)=  \pi^{5/2} (2 M)^{3/2} L_z b^2 T^{1/2}  e^{E/T}
$.
Thus we expect the system to behave as a thermostat.
In our simulation we let this thermostat weakly interact with a
neutral gas of 3 particles confined in a box, and recorded the
probability $p(v)$ to find the absolute value of any of the $3\times 3$
velocity components of the neutral gas at value $v$ during the simulation.
As with the 1D simulation, the 3+1 particles were interacting with each other and with the fixed box
walls via the truncated Lennard-Jones potential, Eq. (\ref{eq:V(x)}).
The logarithmic potential is truncated in the same way as in the 1D case, Eq.
(\ref{eq:phi(x)}), that is we used the potential
\begin{equation}
\varphi(X,Y)=T\ln[(X^2+Y^2+b^2)/b^2]\, .
\label{eq:H-coulomb2D-trunc}
\end{equation}
The results are displayed in Fig. \ref{fig:plot2}.
The truncation of the
logarithmic potential entails a deviation of the density of states from the
exponential form:
$\Omega_{\text{log}}(E)\propto e^{E/T}[\sqrt{\pi} - 2\Gamma(3/2,E/T)]$ where $\Gamma(a,x)$ is the
upper incomplete gamma function.
Note that with $E/T \gg 1$ this deviation vanishes exponentially as 
$\Omega_\text{log}(E) \propto e^{E/T}(\sqrt{\pi}-2\sqrt{E/T}e^{-E/T}$), where we 
have used the asymptotic expansion of the upper incomplete Gamma function \cite{AbramowitzBook}.
This leads to a deviation of the distribution $p(v)$ from
the Maxwellian form. 
For a fixed simulation energy $E_{\text{tot}}$, 
this deviation in $p(v)$ becomes more 
pronounced as the number of degrees of freedom  composing the 
system increases, cf. the inset in Fig. \ref{fig:plot2}. This
can be compensated by increasing the simulation energy $E_{\text{tot}}$.
We estimate that this scales as
$E_\text{tot} \gtrsim c 3NT/2 \simeq c \langle H_S \rangle$, with some constant $c$ 
depending on the required degree of approximation.

\emph{Remarks.--}
Not only can logarithmic potentials be generated artificially, e.g., with 
properly engineered laser fields \cite{Schleich10PRA82}, electrophoretic traps \cite{Cohen05PRL94} or with charged wires, 
but they also 
occur naturally in various situations: 
For example logarithmic potentials govern the motion of stars in elliptic galaxies \cite{Stoica},
determine the interaction of vortices in flow fields \cite{Onsager},
and of probe particles in driven fluids \cite{Levine05EPL70}.
Log-oscillators recently received much attention in regard to their anomalous 
diffusion properties 
\cite{Dechant11JSP145,Dechant11PRL107,Hirschberg11PRE84,Hirschberg12JSM}.
The present work is complementary to these studies 
\cite{Dechant11JSP145,Dechant11PRL107,Hirschberg11PRE84,Hirschberg12JSM}
in the sense that our focus is 
on the dynamics of the particles surrounding the log-oscillator, whereas
their focus is on the dynamics of the log-oscillator itself.

One of the earliest thermostats was proposed by Andersen \cite{Andersen80JCP72}.
In the method of Andersen the system evolves according to
Hamiltonian equations of motion until, at some random time $\tau$,
the velocity of a randomly chosen particle in the system is instantaneously
assigned a new value drawn from a Maxwell distribution with the desired
temperature.
The system then continues it Hamiltonian motion until the next random
event occurs, and so on. Our method can be seen as a fully deterministic
version of Andersen thermostat, where the times at which the collisions
occur and the newly imparted velocities are not drawn randomly,
but follow deterministically from the total system dynamics.

In many studies thermal baths are modeled as
\emph{infinite} collections of  harmonic oscillators or free particles.
In the present  method this infinite collection is replaced by a 
single log-oscillator. It has therefore the evident advantage of not
involving any thermodynamic limit while retaining the
Hamiltonian structure. Roughly speaking, \emph{the thermodynamic
limit is lumped in the singularity of the log-potential}.
At variance with infinite thermal baths whose
temperature is given by the bath's energy per degree of freedom,
log-oscillator  thermostats contain the
temperature as a parameter in the total Hamiltonian. This opens the possibility,
for example, to study the response of a system to a varying temperature,
and take  advantage from the non-equilibrium statistical
mechanical machinery dealing with time dependent Hamiltonians
\cite{Campisi83RMP11}.

Another advantage of our method is that, because the 
Hamiltonian is written in the standard physical system+bath+interaction form: $H=H_S+H_B+h$,
it provides a direct way to control the strength of the interaction $h$, allowing 
also to simulate thermalization to generalized Gibbs states occurring when
the system-bath coupling is not weak \cite{Campisi09PRL102}, which can be a relevant
case for small systems.

\emph{Conclusions.--} We demonstrated that log-oscillators possess infinite 
heat capacity, i.e., they are ideal thermal baths.  As such they have a 
thermostating influence on the dynamics of many-particle systems. 
The resulting deterministic Hamiltonian dynamics are distinct from the Nos\'e-Hoover dynamics.
Unlike previously reported generalized Hamiltonian formulations of Nos\'e-Hoover dynamics,
our Hamiltonian (i) produces thermostated dynamics irrespective of the energy value and
(ii) presents the kinetic terms in standard form. Consequently it is amenable to experimental
realization. Its most promising practical use is as an analog thermostat simulator for the 
experimental investigation of the thermodynamics of small systems, e.g., atomic clusters.

\emph{Acknowledgments.--} The authors thank Sergey Denisov for comments
and Nianbei Li for technical advice. This work was supported by the cluster of excellence
Nanosystems Initiative Munich (P.H.), the Volkswagen
Foundation project No. I/83902 (P.H., M.C.), and the DFG priority program SPP 1243 (P.H., F.Z.).


\end{document}